\documentclass[graybox,envcountchap,natbib,authoryear]{svmult}

\usepackage{mathptmx}       
\usepackage{helvet}         
\usepackage{courier}        
\usepackage{type1cm}        
\usepackage{makeidx}         
\usepackage{graphicx}        
\usepackage{multicol}        
\usepackage[bottom]{footmisc}

\begin{document}

\title*{Gas Accretion and Star Formation Rates$^\dag$} 
\titlerunning{Gas Accretion and Star Formation Rates} 
\author{Jorge S\'anchez Almeida}
\institute{
Jorge S\'anchez Almeida \at Instituto de Astrof\'\i sica de Canarias, La Laguna, Tenerife, Spain, \email{jos@iac.es}\\
Departamento de Astrof\'\i sica, Universidad de La Laguna\\
\\
$^\dag$~To appear in {\bf Gas Accretion onto Galaxies}, eds. A. J. Fox \& R. Dav\'e,
2017, \\Astrophysics and Space Science Library, Springer. 
}
%
%
\maketitle


\abstract{
Cosmological numerical simulations of galaxy evolution show that accretion of metal-poor 
gas from the cosmic web drives the star formation in galaxy disks. Unfortunately, the 
observational support for this theoretical prediction is still indirect, and modeling and analysis 
are required to identify hints as actual signs of star-formation feeding 
from metal-poor gas accretion. Thus, a meticulous interpretation of the observations is crucial, 
and this observational review begins 
with a simple theoretical description of the physical process and the key ingredients 
it involves,  including the properties of the accreted gas and of the star-formation that it induces. 
A number of observations pointing out the connection between metal-poor gas accretion 
and star-formation are analyzed, specifically,
the short gas consumption time-scale compared to the age of the stellar populations,
the fundamental metallicity relationship,
the relationship between disk morphology and gas metallicity,
the existence of metallicity drops in starbursts of star-forming galaxies,
the so-called G dwarf problem,
the existence of a minimum metallicity for the star-forming gas in the local universe,
the origin of the $\alpha$-enhanced gas forming stars in the local universe,
the metallicity of the quiescent BCDs,
and the direct measurements of gas accretion onto galaxies.
A final section discusses intrinsic difficulties to obtain  direct observational 
evidence, and points out alternative observational pathways to further consolidate 
the current ideas.
}

\section{Introduction -- Key physical parameters}\label{sec:intro}

Cosmological numerical simulations show that model galaxies tend to  reach a subtle stationary state where the gas 
accretion rate from the cosmic web balances the star-formation rate (SFR) once outflows are  taken into 
account 
\citep[e.g.,][ see also the contribution by Kere{\v s} in this Book]{2008MNRAS.385.2181F,
2010MNRAS.402.1536S,2012MNRAS.426.2166F,2012MNRAS.421...98D,
2013MNRAS.435..999D,2013MNRAS.433.1425B,2013MNRAS.433.1910F,
2013MNRAS.436.2689A,2014MNRAS.438.1552F}.
The balance is set because the time-scale to transform gas into stars is significantly shorter 
than the Hubble time and, thus, galaxies must rely on external gas accretion to  maintain 
star-formation for a long period of time 
\citep[e.g.,][and also Sect.~\ref{sec:timescale}]{1983ApJ...272...54K,2008AARv..15..189S}.

Numerical simulations reveal an intimate connection between SFR and gas accretion rate, and so,
provide the rationale to study the relation observationally. In order to identify the physical 
parameters  that have to be measured, one can resort to the toy galaxy model often referred to
as bathtub model or self-regulator model. It is amply described in the literature 
\citep[][and also see the contribution by Lilly in this Book]
{1980FCPh....5..287T,1990MNRAS.246..678E,2010ApJ...718.1001B,2011MNRAS.417.2962P,
2012ApJ...750..142B,2012MNRAS.421...98D,2013ApJ...772..119L,2013MNRAS.430.2891D,
2014MNRAS.444.2071D,2014MNRAS.443..168F,2014MNRAS.443.3643P,2015ApJ...800...91H,
2016MNRAS.455.2592R,2015ARAA..53...51S,2015MNRAS.448.2126A}, 
and it provides the physical insight to understand the self-regulation
of the star-formation (SF) process in galaxies. In this simple model, galaxies are described as 
structureless entities characterized by a single mass of gas $M_g$, a SFR, an outflow rate 
$\dot{M}_{out}$, and an inflow rate $\dot{M}_{in}$. 
We take the nomenclature and the equations from the particular implementation by 
\cite{2014AARv..22...71S}. If the model galaxy is isolated and does not receive any
external gas supply, then the initial mass of gas $M_g(0)$ drops exponentially in time $t$
due to star formation (SF),
\begin{equation}
M_g(t)=M_g(0)~{\Large\exp}(-t/\tau_{in}),
\label{scales0}
\end{equation}
with a characteristic time-scale, $\tau_{in}$, given by,  
\begin{equation}
\tau_{in}=\tau_g/(1-R+\eta),
\label{eq:scales1}
\end{equation}
which depends on the so-called gas depletion time-scale $\tau_g$,
\begin{equation}
\tau_g=M_g/{\rm SFR},
\label{eq:kslaw}
\end{equation}
and on the mass loading factor $\eta$, 
\begin{equation}
\eta=\dot{M}_{out}/{\rm SFR},
\label{eq:mass_loading}
\end{equation}
defined to be the scale factor between the SFR and the mass outflows that the SF drives.  $R$ in Eq.~(\ref{eq:scales1}) 
stands for the fraction of stellar mass that returns to the interstellar medium (ISM) rather than being locked into stars 
and stellar remnants. If rather than being isolated our toy galaxy is fed at a gas accretion rate $\dot{M}_{in}(t)$, then
after a transient that lasts $\tau_{in}$, it reaches a stationary state where,
\begin{svgraybox}
\begin{equation}
{\rm SFR}(t)=\dot{M}_{in}(t)/(1-R+\eta).
\label{eq:fundamental}
\end{equation}
\end{svgraybox}
Even if oversimplified, the above equations include all the essential ingredients giving rise to the 
expected relationship between gas accretion rate and SFR. Equation~(\ref{eq:fundamental})  indicates that 
the stationary-state SFR is set by the gas infall rate, becoming zero when the accretion rate goes to zero. 
Often $\eta \gg 1$, and so ${\rm SFR}\ll \dot{M}_{in}$ and $\tau_{in} \ll \tau_g$. In this case, only a minor fraction 
of the accreted  gas is used to form stars.  The rest is returned unused to the circum-galactic medium 
(CGM) and  inter-galactic medium (IGM).  When this happens, the time-scale to consume the gas $\tau_{in}$ becomes much 
shorter than the  already short gas depletion time-scale (Sect.~\ref{sec:timescale}).

Therefore, in order to provide an observational overview of the relation
between gas infall rate and SFR, it is essential to keep in mind and constrain 
the key parameters characterizing the relation, namely, the 
gas depletion time-scale,  the mass loading factor, and the returned mass fraction. 
Thus, the first section of the paper collects observational constraints on these parameters
(Sect.~\ref{sec:phyparam}). Section~\ref{sec:evidence} constitutes the main 
body of the work, and it describes observational evidence for a relationship 
between SFR and metal-poor gas accretion.
Unfortunately, despite the large volume of circumstantial evidence for  
feeding from external gas accretion,  we still lack direct evidence.
Several factors explain the  difficulty.  They are pointed out and discussed in 
Sects.~\ref{sec:complications} and \ref{sec:conclusions}, where we also
mention future lines of research.


\section{Characteristic physical parameters}\label{sec:phyparam}

\begin{figure}
\begin{center}
\includegraphics[width=3.0in]{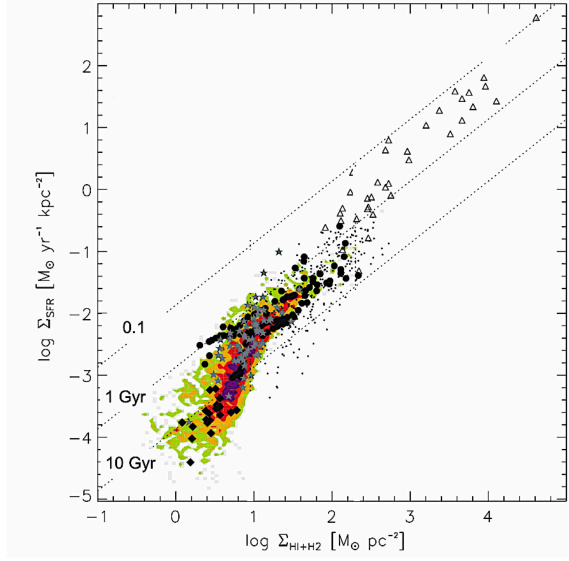}
\end{center}
\caption{
Kennicutt-Schmidt
relation at sub-kpc scales, i.e., SFR surface density versus 
gas surface density (labeled in the plot as $\Sigma_{{\rm HI+H2}} = \Sigma_g$).
The slanted dotted lines represent the expected relation for constant values of $\tau_g$ of 
0.1, 1, and 10~Gyr. They encompass the 
observed data points. Adapted from \citet[][their Fig.~15]{2008AJ....136.2846B}.
}
\label{fig:ks}
\end{figure}

{\bf Gas depletion time-scale $\tau_g$.}
The ratio between $M_g$ and SFR, $\tau_g$, can be measured directly.  
The mass of gas and the SFR are correlated as given by the so-called
Kennicutt-Schmidt  relation  \citep{1959ApJ...129..243S, 1998ApJ...498..541K}, 
which is usually formulated  in terms of the gas surface density, $\Sigma_g$, 
and the SFR surface density,  $\Sigma_{\rm SFR}$,
\begin{equation}
\Sigma_{\rm SFR} = A\,\Sigma_g^N,
\label{eq:k-s}
\end{equation}
so that 
\begin{equation}
\tau_g=\Sigma_g/\Sigma_{\rm SFR} = A^{-1}\,\Sigma_g^{(1-N)}.
\end{equation}
\citet{2012ARAA..50..531K} give A and N for disk-averaged galaxies, 
which  lead to $\tau_g$ from 3 to 0.5~Gyr for gas surface densities between 
10 and 1000~$M_\odot\,{\rm pc^{-2}}$, typical of star-forming galaxies.  
Similar $\tau_g$ are obtained when considering spatially resolved 
measurements with sub-kpc resolution; see Fig.~\ref{fig:ks}, 
adapted from \citet[][]{2008AJ....136.2846B}. It shows the lines of constant 
$\tau_g$, and how the observations in the range of interests are close to the $\tau_g=1$\,Gyr line.
The Kennicutt-Schmidt  law provides an empirical recipe linking SF  with its fuel, 
and it has become central to the current numerical simulations of galaxy formation
\citep[e.g.,][]{2008MNRAS.383.1210S,2008MNRAS.387.1431D,2012MNRAS.426..140D}. 
Therefore, the relation has deserved much attention in the literature, rendering 
depletion time-scales in the range of the above values 
\citep[see][and references therein]{2016ApJ...820..109F}. 

Observations suggest that $\tau_g$ decreases with increasing redshift, so that it goes from
 0.5 to 2~Gyr for galaxies in the redshift range between 2 and 0 
\citep[e.g.,][]{2010MNRAS.407.2091G,2014ApJ...787L...7G}. This is consistent with the theoretical
expectation that $\tau_g$ scales with the instantaneous Hubble time, 
$t_H$,  as $\tau_g \simeq  0.17\,t_H$, so that at any time $\tau_g\ll t_H $ \citep{2013MNRAS.435..999D}.

{\bf Mass loading factor $\eta$.}
Stellar feedback is an essential ingredient of the current galaxy formation scenario
since is it partly controls the predicted number of low-mass galaxies. Thus, its tuning 
allows us to reproduce the observed galaxy mass function 
\citep[e.g.,][]{2012RAA....12..917S,2014MNRAS.444.1518V,2015MNRAS.446..521S}.
The mass loading factor quantifies the importance of this stellar feedback 
 (Eq.~[\ref{eq:mass_loading}]). Using the toy model, it is easy to argue that
$w$ should increase with decreasing stellar mass $M_\star$, so that for 
low-mass galaxies,   
\begin{equation}
\eta\geq 1.
\label{eq:condition_w}
\end{equation}
In the stationary state, the metallicity of the gas  that forms  stars, $Z$, is independent of SFR and $M_g$.
$Z$ only depends on parameters characteristic of
the stellar population, and on $\eta$, explicitly,
\begin{equation}
Z=Z_{in}+y\,(1-R)\big/(1-R+\eta),
\label{eq:metallicity}
\end{equation} 
with $Z_{in}$ the metallicity of the accreted gas, and $y$ the stellar yield (the mass of ejected metals
per unit mass locked into stars). Galaxies follow a well known mass-metallicity relation
\citep[MZR; see, e.g.,][]{1989ApJ...347..875S,2004ApJ...613..898T} 
which, according to Eq.~(\ref{eq:metallicity}), can only be 
due to the dependence of $\eta$ on $M_\star$ since $y$ and $R$ are universals given 
the initial mass function (IMF). 
Moreover, for $\eta$ to modify $Z$ in a substantial way,
Eq.~(\ref{eq:condition_w}) must be satisfied. 

The fact that $\eta$ varies with $M_\star$ reaching large values for low-mass galaxies 
is found both in numerical simulations and in observational works.  In order to reproduce the 
MZR,  \citet{2011MNRAS.417.2962P}, \citet{2012MNRAS.421...98D},
and \citet{2013MNRAS.430.2891D}
use $\eta$ varying from 1 to 6 when $M_\star$ goes from $10^{11}$ to $10^9\,M_\odot$.
The numerical simulations by \citet{2012ApJ...760...50S} lead to $\eta$ between 1 and 10 for   
$M_\star$ from $2\times 10^{11}$ to $10^9\,M_\odot$. In the cases modeled
by \citet{2016ApJ...824...57C}, it goes from 0.5 to 50 for galaxies 
with virial masses from $10^{12}$ to $2\times 10^9\,M_\odot$.
The numerical simulations carried out by \citet{2013MNRAS.434.2645D} assume $\eta$ to be 
proportional to $M_\star^{-1/3}$, with an even steeper dependence at low masses.
The numerical simulations of giant star-forming clumps in gas-rich galaxies by 
\citet{2014ApJ...780...57B} render an effective  $\eta$ exceeding 2. 
\citet{2016MNRAS.455..334T} model supersonic-turbulence driven outflows, and 
infer $w$ ranging from $10^{-3}$ to 10.
Observations of Mg\,II absorption around massive galaxies with $M_\star\sim 10^{11}\,M_\odot$
are used by \citet{2014ApJ...794..130B} to set a conservative lower limit to $\eta$, which 
has to be larger then 0.24. Mass loading factors observed at
high redshift  generally refer to massive objects ($M_\star >  10^{10}\,M_\odot$), and turn out to be 
between 0.5 and 2 \citep{2012ApJ...761...43N,2012ApJ...760..127M}. It is not uncommon to infer
factors up to 10 in local dwarfs \citep{1999ApJ...513..156M,2005ARAA..43..769V}. An extreme case is
 presented by  \citet{delolmo16}, where faint multiple components in the wings of H$\alpha$
are interpreted as produced by SN driven outflows with $\eta$ often larger than 20.

{\em\bf Returned fraction $R$}.\label{subsec:fraction}
Once the IMF is set, the stellar mass returned to the ISM is provided
by stellar evolution models. 10 Gyr after the starburst, $R$ is typically in the range between 
0.2 and 0.3 when a  Salpeter IMF is adopted \citep[e.g.,][]{2005AG....46d..12E,2011ApJ...734...48L}. 
It can reach 0.5 for a Kroupa IMF, that has more massive stars for the 
same total mass of the stellar population \citep[e.g.,][]{2011ApJ...734...48L,2016MNRAS.456.1235S}.    

\section{Evidence for a relationship between the SFR and the gas infall rate}\label{sec:evidence}


\subsection{The gas-consumption timescale}\label{sec:timescale}

The  time-scale to consume the gas is too short for the observed stellar ages, therefore,
a continuous gas supply is needed to explain why all types of galaxies have been forming 
stars for extended periods of time.  This is an old idea 
\citep[e.g.,][]{1963ARAA...1..149R,1983ApJ...272...54K} that 
has been reformulated in various ways along the years. A updated account is given next.

\begin{figure}
\begin{center}
\includegraphics[width=0.6\textwidth]{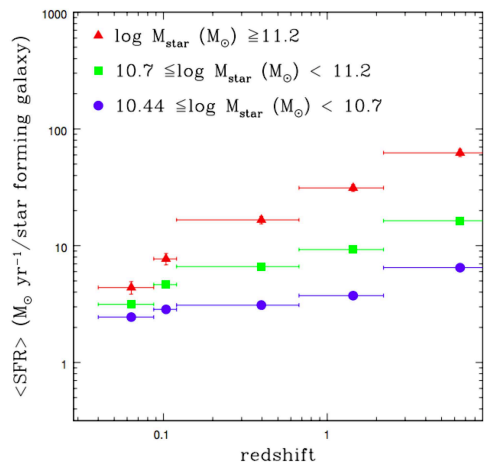}
\end{center}
\caption{
Observed  SFR versus redshift for typical galaxies forming stars at present. The  
galaxies are divided into  three bins according to the present stellar mass (see the inset).
Galaxies of all masses have been forming stars over the whole history of the Universe.
Figure adapted from \citet[][their Fig.~6]{2015MNRAS.450.2749G}.
}
\label{fig:sfr_time}
\end{figure}
Except for the most massive ones, galaxies have been forming star
during the whole life-time of the universe. Figure~\ref{fig:sfr_time} shows the observed SFR versus 
redshift for typical disk galaxies forming stars at present. The galaxies have been divided into  three 
bins according  to the present stellar mass. The figure shows how their SFR has been declining 
slightly with time but, overall, galaxies of all masses maintain a significant level of SF along the Hubble 
time \citep[see also][]{2004Natur.428..625H}. If no gas supply is provided, the SF shuts off exponentially 
with a time-scale of 
\begin{equation}
\tau_{in}\simeq\,0.75\,{\rm Gyr}, 
\end{equation}
where we have plugged into Eq.~(\ref{eq:scales1})  the values 2\,Gyr, 2, and 0.3 for $\tau_g$, $\eta$ and $R$,
respectively (Sect.~\ref{sec:phyparam}). Since the SF has been active for much longer, a supply of external
gas must have been feeding the SF process in all galaxies.

\subsection{Relationship between stellar mass, SFR, and gas metallicity}

\begin{figure}
\begin{center}
\includegraphics[width=0.6\textwidth]{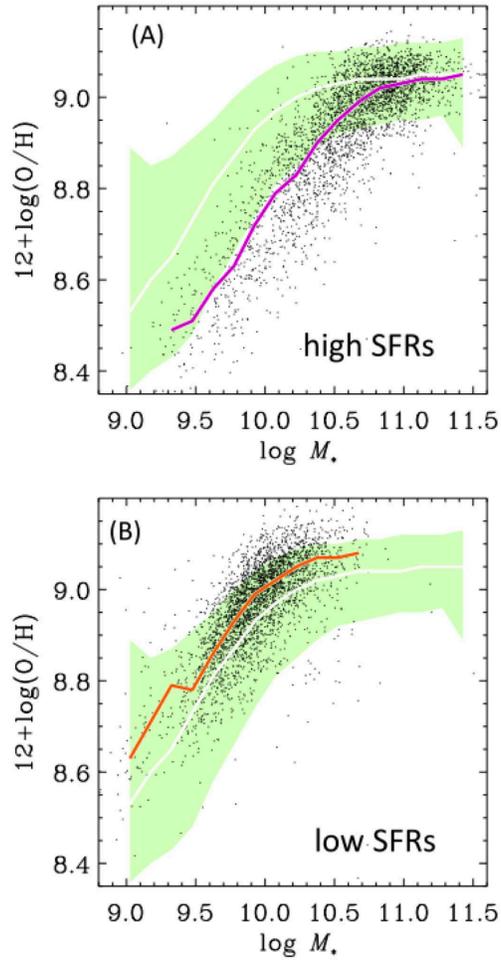}
\end{center}
\caption{
Gas-phase metallicity versus  stellar mass for galaxies with extreme SFR. 
From a sample of  $\sim 250,000$ SDSS objects,  the dots in the upper panel (A) show the galaxies 
with the highest SFRs. The lower panel (B) contains those
galaxies with the lowest SFRs. The colored lines represent the 
median of the distribution of points for a given stellar mass. 
The green  shaded region is the same
in (A) and (B) and gives the 90 percentile range of the full sample, with
the white line representing the median. Note the systematically lower metallicity
for the sample of higher SFR.  Figure adapted from \citet{2014ApJ...797..126S}.}
\label{fig:fmr}
\end{figure}

In two independent papers, \citet{2010MNRAS.408.2115M} and \citet{2010AA...521L..53L} found that the 
scatter in the well-known mass-metallicity relation correlates with the SFR.
Such correlation had been suggested in previous studies 
\citep{2008ApJ...672L.107E,2009ApJ...695..259P,2010AA...521A..63L}.
The fact that galaxies with higher SFR  show lower metallicity at a given 
stellar mass  is called fundamental metallicity relation (FMR).  
Figure~\ref{fig:fmr} shows a recent account of the FMR by \citet{2014ApJ...797..126S}.
Galaxies are separated into two plots according to their SFR, and the difference is clear: 
high SFR objects (panel A) have lower metallicity than low SFR objects (panel B).  

\begin{figure}
\includegraphics[width=0.9\textwidth]{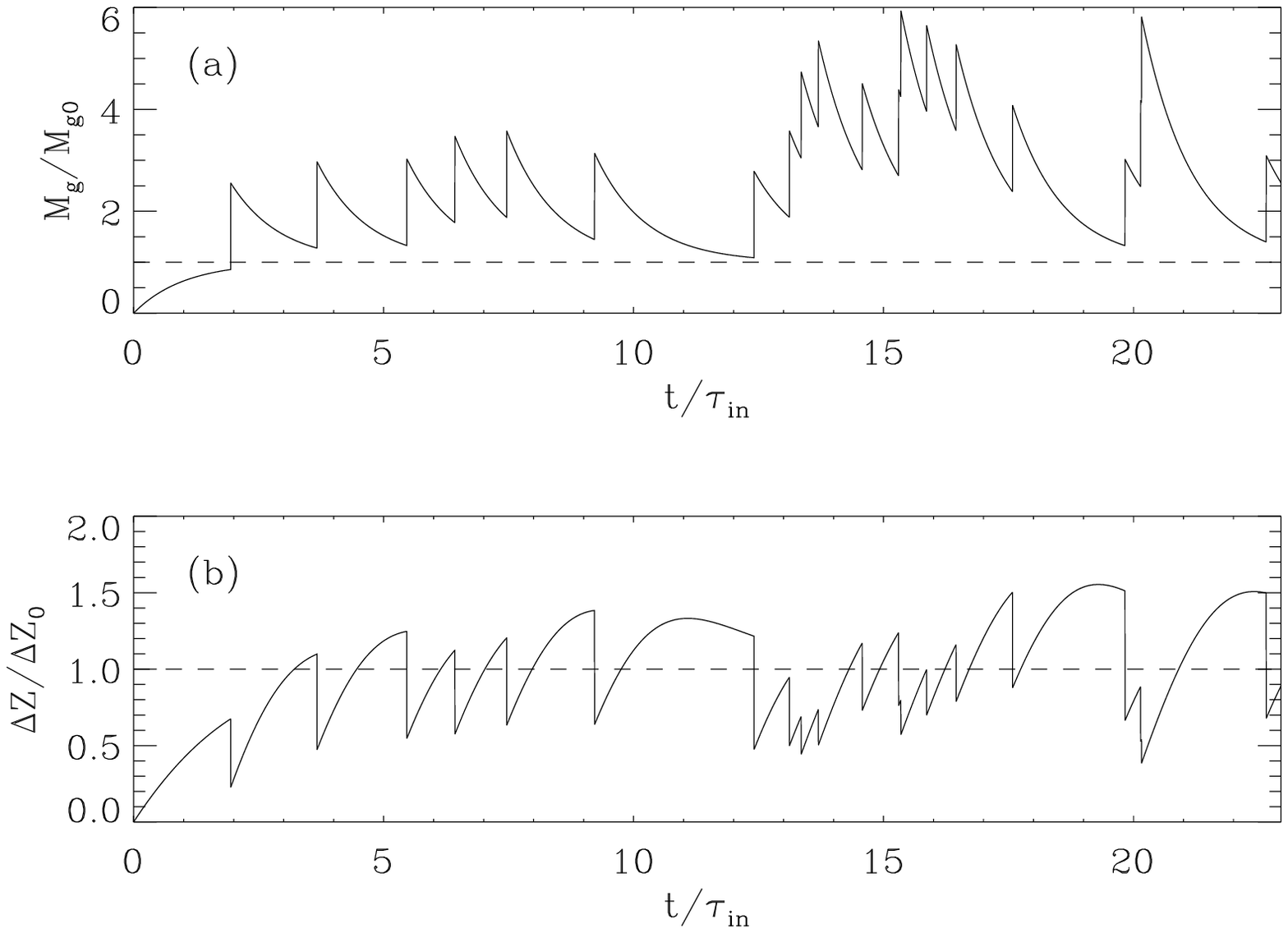}
\caption{
Variation with time of gas mass (a)  and metallicity (b) in a star-forming system where gas 
clumps  are accreted randomly, i.e., at random times and with random masses. Time
units are normalized to $\tau_{in}$, the characteristic time-scale for the exponential
fall-off of the gas content. The gas of mass and metallicity are normalized to 
their stationary-state value, indicated by horizontal dashed lines in the figures.
}
\label{fig:gasz}
\end{figure}

Neither metal-rich outflows nor the variation of the SF efficiency with $M_\star$ explain the 
FMR. However, the  stochastic feeding of the SF process with external metal-poor gas 
provides a natural explanation. The advent of external gas does not change $M_\star$, but 
it decreases the mean metallicity of the star-forming gas while simultaneously triggering SF. 
As time goes on, the SF consumes gas and increases its metallicity,  until new metal-poor gas 
arrives and the cycle re-starts. The process is illustrated in 
Fig.~\ref{fig:gasz}. It contains the temporal variation of the gas mass and 
metallicity predicted by the toy model in Sect.~\ref{sec:intro}, assuming the gas accretion 
events to be stochastic.  
The figure shows how the arrival of gas at a particular time increases the 
gas mass and thus the SFR (Eq.~[\ref{eq:kslaw}]). This gas accretion event comes with a
drop in metallicity. The gas is consumed by the SF, that rises the metallicity in a process
that in the long run yields  the stationary-state value given in Eq.~(\ref{eq:metallicity}).     
If a collection of galaxies with similar stellar masses are observed at different phases of 
the cycle triggered by a gas accretion event, they will show a dispersion in metallicity
which anti-correlates with the instantaneous SFR, i.e., they will show the FMR. 
This explanation was already advanced by \citet{2010MNRAS.408.2115M} and 
it is generally accepted today. 

The FMR has received considerable attention in the literature\footnote{The discovery papers have about 600 citations in the ADS.}, 
both from an observational point of view, and from the point of view of its interpretation. 
Sometimes the claims seem to be contradictory, although the community 
is reaching a consensus in the sense that (1) the FMR is not an observational artifact, 
(2) it changes with redshift so that all metallicities decrease with increasing redshift, 
and (3) it is produced by metal-poor gas accretion triggering SF. The next
paragraphs summarize the recent observational and  theoretical  work on the 
subject.

%
{\bf Observations of the FMR.} 
The correlation between SFR and metallicity is weaker if systematic errors are taken into account 
\citep{2016ApJ...827...35T}. The correlation may disappear depending on the strong-line ratio
used to estimate metallicities \citep{2016ApJ...823L..24K}.  The FMR remains even if different metallicity 
and SFR indicators are used \citep{2014ApJ...797..126S,2013ApJ...765..140A}. 

The FMR disappears when using single H\,II regions rather than galaxy 
integrated parameters \citep{2013AA...554A..58S}. There is no FMR in the local 
star-forming  galaxies analyzed by \citet{2014AA...561A..33I}, whereas it is present 
in the local sample discussed by \citet{2016MNRAS.457.2929W}. Arguments against the existence 
of a FMR in star-forming galaxies with redshift smaller than one are presented by 
\citet{2015MNRAS.451.2251I}. There is no
obvious FMR for galaxies with redshifts between 1 and 2, according to \citet{2014AA...569A..64D}.
There is not relationship at redshift 0.8 \citep{2015AJ....149...79D}.

At redshift around 1.4, the deviation from the MZR depends on the SFR, 
so that galaxies with higher SFR  show lower metallicities at a given $M_\star$
\citep{2012PASJ...64...60Y}.
The FMR is still in place at redshift around 0.9, but the metallicities are systematically lower
given $M_\star$ and SFR \citep{2014ApJ...780..122L,2015ApJ...805...45L}. 
Galaxies with younger and more vigorous star formation 
tend to be more metal poor at a given $M_\star$
\citep[redshift between 0.3 and 0.9;][]{2014AA...568L...8A}.
The FMR is in place at $z \geq 2$ \citep{2014MNRAS.440.2300C}.
Galaxies at redshift 2 show  evidence that the SFR is still a second parameter in the 
MZR, and are consistent with a non-evolving FMR \citep{2014ApJ...792....3M}. 
The FMR is in place at redshift 1.6, but it has evolved with respect to the FMR in the local universe
so that metallicities are smaller \citep[][]{2014ApJ...792...75Z}.
At redshift 3.4, the  metallicity generally anti-correlates with the distribution of SFR  and with 
the gas surface density, although the relation differs from the FMR in the local universe 
\citep{2014AA...563A..58T}.
The evolution of the FMR previously reported in the literature may be an artifact 
introduced by the use of the different metallicity indicators at different redshifts 
\citep{2014MNRAS.440.2300C}. 
There is a FMR at redshift 0.7 that seems to agree with the local one \citep{2015AA...577A..14M}.
There is no correlation at redshift 2.3 \citep{2015ApJ...799..138S}.
The FMR evolves with redshift \citep{2016MNRAS.458.1529B}.
There is no significant dependence of the metallicity on SFR at fixed redshift and $M_\star$ 
\citep[objects with redshifts between 0.6 and 2.7; ][]{2016ApJ...827...74W}.

There is also a {\em more fundamental} FMR where the SFR is replaced with the gas
 mass  \citep{2013MNRAS.433.1425B}.  The scatter of the FMR is reduced if H\,I mass is used instead of SFR 
\citep{2015ApJ...812...98J}. The central role assigned to the gas mass  at the sacrifice of the SFR is also 
defended by  \citet{2016arXiv160604102B} and \citet{2016MNRAS.455.1156B}.
Moreover, \citet{2015MNRAS.446.1449L} claim that stellar age,  rather than SFR or gas mass, 
is the third parameter in the FMR .

There is a correlation between the metallicity gradient along the 
radial distance in a galaxy and the SFR, in the sense that galaxies with high SFR tend to 
show flatter gradient \citep{2014MNRAS.443.2695S}.

{\bf Interpretations of the FMR.}
Most of the available explanations are based on simple analytical models  very much in the spirit of
the one  described in Sect.~\ref{sec:intro}. For example, \citet{2013ApJ...772..119L} 
present a model galaxy whose properties self-regulate due to the short gas depletion time-scale. 
The model galaxy is near the stationary state, but the gas reservoir available to form stars is allowed to change 
in time. This drives the system out of the stationary state and provides a dependence of the metallicity on the 
SFR and mass gas.  (The metallicity does not depend on the SFR in the stationary state; see Eq.~[\ref{eq:metallicity}].)
The work by \citeauthor{2013ApJ...772..119L} successfully reproduce the FMR, allowing both $w$ and 
$\tau_g$ to vary with stellar mass. It reproduces the overall drop of metallicity with increasing
redshift by steadily increasing the gas infall rate. 
Other works with this type of interpretation are those by 
\citet{2011MNRAS.416.1354D}, \citet{2012ApJ...750..142B},
\citet{2013MNRAS.430.2891D}, \citet{2014MNRAS.443..168F}, \citet{2014MNRAS.441.1444P},
and \citet{2015ApJ...800...91H}.  
Mergers are also able to reproduce the FMR according to \citet{2015MNRAS.451.4005G}.

\citet{2014MNRAS.444.2071D} use one of these simple toy models to study the redshift 
dependence of the FMR, finding problems to reproduce some of the observational constrains, in particular,
the ratio SFR/$M_g$.  The need to go beyond simple models because they do not reproduce 
the observed variation with redshift of SFR/$M_g$ is also argued by \citet{2014MNRAS.443.3643P}.

\citet{2012MNRAS.422..215Y} use thousands of galaxies from  dark-matter numerical simulations to interpret the FMR. 
Baryons that follow the dark-matter are added, generating a non-stationary clumpy gas accretion that
drives the evolution of the model galaxies. The numerical simulation reproduces the main observational trends, 
including an apparent turnover 
of the mass-metallicity relationship at  very high $M_\star$. The temporal evolution of the 
gas mass and metallicity of individual galaxies is qualitatively similar to the variations displayed
in Fig.~\ref{fig:gasz}. 

\citet{2013ApJ...770..155R} present SPH-cosmological simulations of hundreds of 
galaxies. Surprisingly,  more active galaxies in terms of  SFR are also metal-richer (see
their Fig.~12).  The reason of this contradictory result is not properly understood. 

\citet{2015MNRAS.452..486D} employ hydrodynamical zoom-in cosmological 
simulations of 500 galaxies to study the scaling relations.  The model galaxies show 
the trends corresponding to the observed FMR.  They also find that satellite galaxies 
have higher metallicity for the same stellar mass, as it is indeed observed \citep{2012MNRAS.425..273P}. 

\citet{2016ApJ...826L..11K} find galaxies at redshift around 2 following the FMR. 
They show that the gas masses and metallicities required to reproduce the observed FMR
are consistent with cold-accretion predictions obtained from their hydrodynamical simulations.


\subsection{Relationship between lopsidedness and metallicity}\label{sec:lopsided}

\begin{figure}
\includegraphics[width=1.0\textwidth]{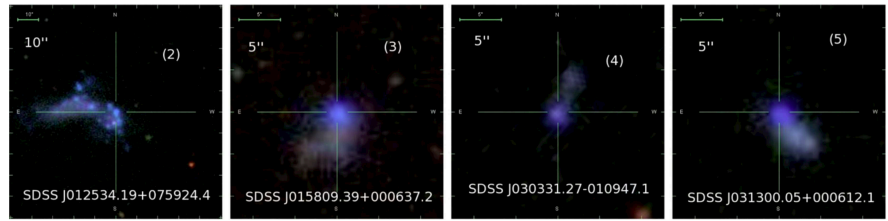}
\caption{Typical set of galaxies selected only because they are XMP.
Surprisingly, they tend to show cometary or tadpole morphology, 
with a bright blue head and a faint redder tail. The images are composite color images 
from SDSS  broad-band filters,  therefore, they trace stellar light. 
Adapted from Fig.~5 in \citet{2011ApJ...743...77M}.}
\label{fig:xmps}
\end{figure}

Surprisingly, the extremely metal poor (XMP) galaxies of the local universe turn out to show
a particular morphology consisting of a bright head and a faint tail, which is commonly referred 
to  as {\em cometary} or {\em tadpole}. This correspondence between low metallicity and shape
was first noted by \citet{2008AA...491..113P},
and then it has been confirmed in other studies 
\citep[e.g.,][]{2011ApJ...743...77M,2013AA...558A..18F,2016ApJ...819..110S}.
These morphologies represent 80\,\%\ of the objects in the XMP catalog used by 
\citet{2013AA...558A..18F}. The tadpole morphology is not unusual at high redshift,
where galaxies tend to be clumpy and elongated \citep{2005ApJ...631...85E}, however it is
rather uncommon in the local universe where XMPs reside. For reference, only 0.2\,\%\ of 
the star-forming galaxies in the Kiso survey are cometary \citep{2012ApJ...750...95E}.
Figure~\ref{fig:xmps} displays several of these XMP galaxies with the characteristic 
morphology.

Even though XMP galaxies are a very particular type of galaxy, the morphology-metallicity relation
that they exhibit is only the extreme case of a common behavior followed by many star-forming galaxies.   
\citet{2009ApJ...691.1005R} quantify the lopsidedness of 25,000  star-forming galaxies from SDSS  using 
the amplitude of the $m = 1$ azimuthal Fourier mode. At a fixed mass, the more metal-poor galaxies turn
out to be more lopsided, extending the morphology-metallicity relation to the full population
of star-forming galaxies. 

This property of the XMPs and the other galaxies can be naturally understood within 
the {\em gas accretion triggering} scenario. The actual characteristics of the starbursts 
induced by gas accretion are far from being properly understood and modeled 
\citep[e.g.,][see also Sect.~\ref{sec:complications}]{2014MNRAS.442.1830V,2006PASP..118..833C}. 
However, a  few general trends  seem to be clear. 
The accreted gas is metal-poor \citep[e.g.,][]{2009Natur.457..451D,2012MNRAS.423.2991V},
and it induces off-center giant star-forming clumps that gradually migrate 
toward the disk centers \citep{2010MNRAS.404.2151C,2014MNRAS.443.3675M}.
The giant star-forming clumps may be born {\em in-situ} or {\em ex-situ}. In the first
case, the accreted gas builds up the gas reservoir in the disk to a point where disk instabilities 
set in and trigger SF. In the second case, already formed clumps are 
incorporated into the disk. They may come with stars and dark matter, and thus, 
they are often indistinguishable from gas-rich minor mergers \citep{2014MNRAS.443.3675M}.
In any case, a significant part of the SF in the disks occurs in these giant clumps.    
As a result of the whole process, the gas accretion produces bright off-center starbursts 
increasing the lopsidedness of the host disk. This increase and the decrease of 
metallicity come hand-to-hand together, giving rise to a relation between 
morphology and metallicity qualitatively similar to the observed one.  

We note, however, that the same trend can also be reproduced by 
gas-rich metal-poor mergers \citep[e.g.][]{2009ApJ...700.1896K,2016MNRAS.456.3032P}. 
As we pointed out above, {\em gas-rich minor mergers} and {\em gas accretion events} 
are often impossible to distinguish, both observationally and from the point of view of the
numerical  simulations. On the one hand, it is unclear how to define {\em galaxy} at the 
low-mass end of the galaxy mass function. If the presence of stars is essential 
\citep[see][]{2011PASA...28...77F}, whether a gas dominated system is or is not a galaxy 
ultimately depends on the sensitivity of the observation
\citep[e.g.,][]{2014ApJ...787L...1C,2015MNRAS.452.2680S,2015ApJ...801...96J}. 
On the other hand, the presence or absence of stars in a particular dark-matter halo
of a numerical simulation depends on details of the assumed sub-grid physics,  
which  may or may not be adequate to describe the formation of stellar systems in 
objects with sizes and masses at the resolution of the simulation.

This remarkable association between SFR and lopsidedness has been observed in the H\,I morphology too.
\citet{2014MNRAS.445.1694L} find that  dwarfs with active SF have more asymmetric outer H I envelopes 
than typical irregulars. Moreover, galaxies hosting an old burst ($\ge\,100$\,Myr) have more symmetric H\,I 
morphology than those with a young one ($\le\,100$\,Myr).

\subsection{Metallicity drops in starbursts of local star-forming galaxies}\label{sec:drops}

According to the conventional wisdom, the gas of the local gas-rich dwarf galaxies 
has uniform metallicity  \citep{1996ApJ...471..211K,2009ApJ...705..723C,2015MNRAS.450.3254P}. 
However, there is mounting evidence that some particular objects do show metallicity inhomogeneities 
\citep[][]{2006AA...454..119P,2009AA...503...61I,2010ApJ...715..656W,
2011ApJ...739...23L,2012AA...546A.122I,2013ApJ...765...66H,2013ApJ...767...74S,
2014MNRAS.441.2034T,2014ApJ...783...45S,2014MNRAS.445.1104R}.
They are often associated with  star-forming regions, so that a drop in metallicity occurs in 
regions of intense  SF. 

\begin{figure}
\includegraphics[width=0.50\textwidth]{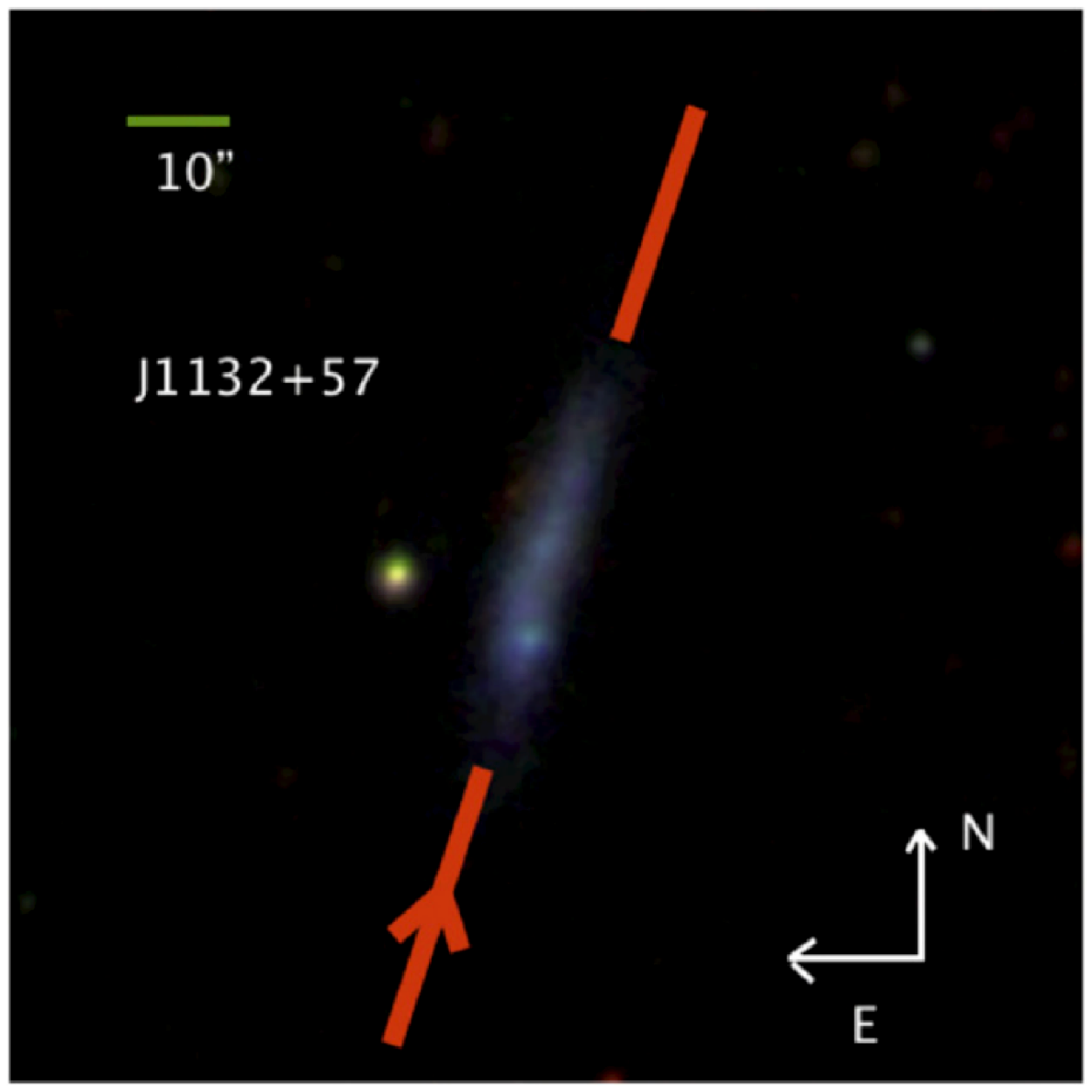}
\includegraphics[width=0.50\textwidth]{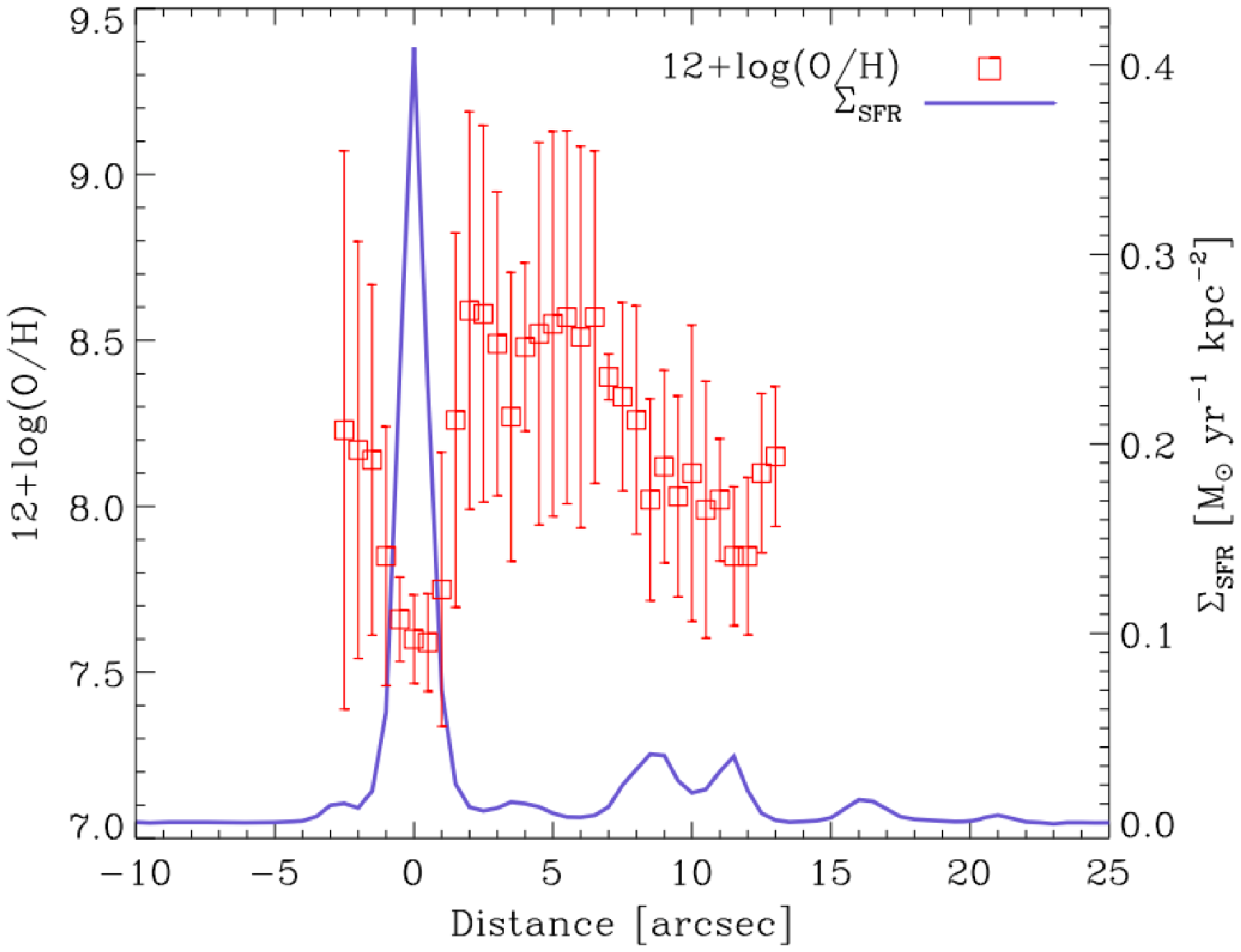}
\caption{
Left: 
SDSS image of the XMP galaxy  J1132+57, with the red bar indicating the position 
of the slit during observation. The arrows indicate north and east, and the scale on the top left corner
corresponds to 10 arcsec.
Right: 
variation of  surface SFR (blue solid line) and oxygen abundance 
(red symbols with error bars) along the slit. Note the drop in abundance associated with the peak 
SFR. Figure adapted from \citet{2015ApJ...810L..15S}.}
\label{fig:sa15a}
\end{figure}
\begin{figure}
\begin{center}
\includegraphics[width=4.0in]{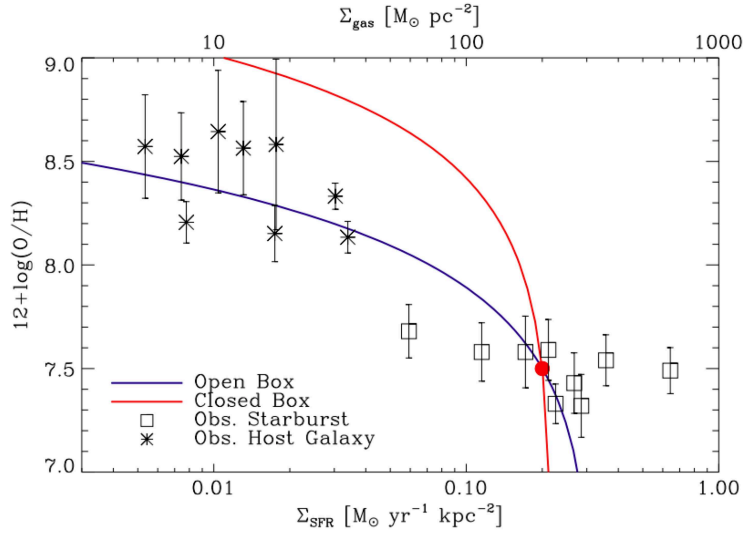}
\end{center}
\caption{Summary plot with the oxygen abundance of the starburst (square symbols) 
and the host galaxy (asterisks) for the XMP galaxies studied by \citet{2015ApJ...810L..15S}. 
They are represented versus the surface SFR inferred from H$\alpha$. The
star-forming clumps are 0.5 dex metal-poorer than the host, and have a SFR between
10 and 20 times larger. The axis on top gives the gas surface density. 
The lines show the chemical evolution of a clump at the position of the red bullet
depending on whether it evolves as a closed-box (red line) or as an open box (blue line).
Taken from \citet{2015ApJ...810L..15S}.
}
\label{fig:sa15b}
\end{figure}

\citet{2015ApJ...810L..15S} carried out a systematic study of the variation of gas metallicity along the 
major axis of a representative sample of XMP galaxies. Metallicities were inferred
using HCM \citep{2014MNRAS.441.2663P}, a code that compares the observed optical emission 
lines with photoionization models and which provides metallicity measurements in agreement 
with direct-method within  0.07\,dex. Figure~\ref{fig:sa15a} contains the result for one of the galaxies. It shows a
clear drop in metallicity at the peak surface SFR. The pattern is the same 
in 9 out of the 10 studied galaxies. The XMP star-forming clumps are immersed 
in a host galaxy which is several times more metal-rich.  
Figure~\ref{fig:sa15b} summarizes these results. 
Independent observations proof that the XMP galaxies rotate, 
and that the star-forming clumps of low metallicity are dynamically 
decoupled from the underlying disk \citep{delolmo16}.

\begin{figure}
\begin{center}
\includegraphics[width=3.0in]{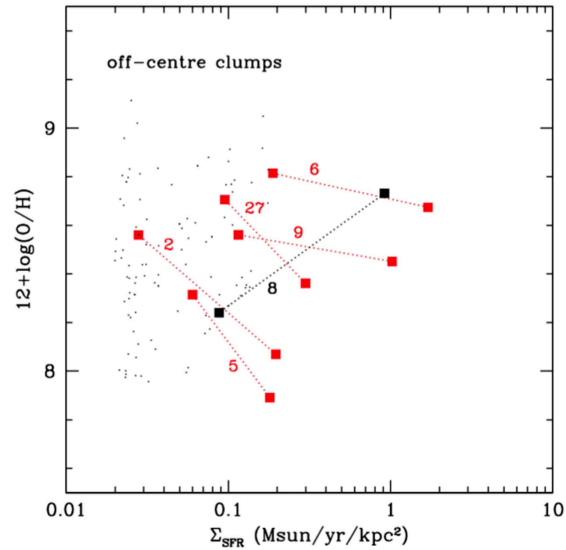}
\end{center}
\caption{
Metallicity [$12+\log({\rm O/H})$] versus surface SFR ($\Sigma_{\rm SFR}$)
for a number of clumps and their nearby intra-clump medium in 
the cosmological numerical simulations of galaxies by 
\citet{2016MNRAS.457.2605C}. Each pair is joined by a dotted line.
In all but one case (the one shown in black), the region of large
SFR (the clump) has lower metallicity than the nearby region of small
SFR (the intra-clump medium).   
The small black dots represent 100 randomly chosen apertures in one 
of the model galaxies.
Adapted from Fig.~3 in \citet{2016MNRAS.457.2605C}.}
\label{fig:ceverino}
\end{figure}

The existence of localized metallicity drops suggests a recent metal-poor gas accretion episode.  
The time-scale for the azimuthal mixing of the gas in turbulent disk  is short, 
of the order of a fraction of the  rotational period \citep[e.g.,][]{2002ApJ...581.1047D,2012ApJ...758...48Y,2015MNRAS.449.2588P},
equivalent to a few hundred Myr. 
Therefore, the metal-poor gas forming stars must have arrived to the disk recently, as naively 
expected for SF episodes driven by external metal-poor gas accretion. 
As we discussed in Sect.~\ref{sec:lopsided}, the triggering of SF feeding from external gas 
is a complex process not properly understood yet (see also Sect.~\ref{sec:complications}). 
There is a significant
degree of gas mixing in the CGM (Sect.~\ref{sec:complications}) and the {\em naive} interpretation may not be correct. 
However, the cosmological numerical simulations of galaxies  analyzed by 
\citet{2016MNRAS.457.2605C} are reassuring. The model galaxies produce off-center star-forming 
clumps with a metallicity lower than the metallicity of their immediate surroundings. 
Figure~\ref{fig:ceverino}, taken from \citet{2016MNRAS.457.2605C}, shows metallicity 
versus surface SFR for a number of clump intra-clump pairs. Each pair 
is joined by a dotted line. They follow a clear pattern where the point of 
lower metallicity coincides with the point of larger SFR. Qualitatively, the 
figure resembles the behavior of the star-forming clumps in 
XMPs (see Fig.~\ref{fig:sa15b}).
 

\subsection{The traditional G-dwarf problem}

This observation is included here both for historical reasons, and because the accepted 
interpretation is easy to understand in terms of the toy model described 
in Sect.~\ref{sec:intro}. The so-called {\em G-dwarf problem } was the first clear indication that
an external metal-poor gas supply was needed to explain the observed properties a 
stellar population.

\begin{figure}
\begin{center}
\includegraphics[width=3.5in]{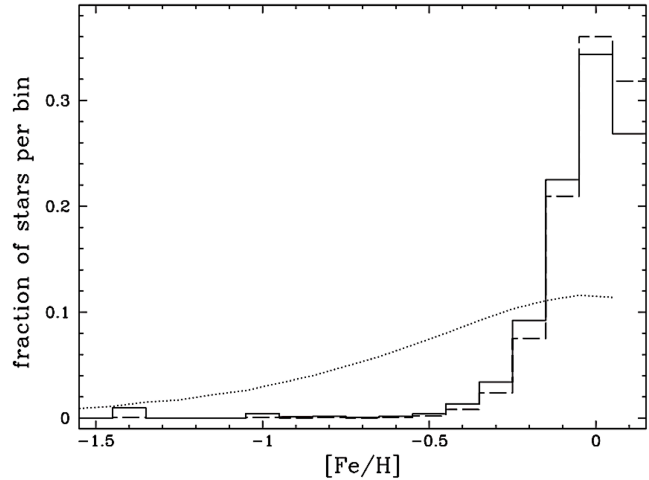}
\end{center}
\caption{Fraction of stars in each 0.1 dex metallicity bin.
[Fe/H] denotes the Fe abundance referred to the solar metallicity in a decimal  
logarithm scale.  The dashed and solid lines represent observed uncorrected and 
corrected data, respectively. The dotted line is the distribution predicted
by a closed-box chemical evolution model, and it largely deviates from the observed 
one. The observed distribution, from \citet{2012MNRAS.422.1489W}, corresponds to M dwarf stars in the
solar neighborhood, but is very similar to the  distribution of G dwarfs discussed in 
the text  \citep[see, e.g.,][]{1996MNRAS.279..447R}.
}
\label{fig:mdwarf}
\end{figure}
If a system of pure metal-poor gas evolves as a closed box, each new generation of 
stars must be less numerous and more metal-rich than the previous one. Therefore, in such 
a system, the number of stars is expected to decrease with increasing metallicity.
However, the distribution of metallicities of the G dwarf stars in the solar 
neighborhood does not show the fall-off expected in a closed box. There is a deficit of 
sub-solar metallicity G dwarf stars in the solar neighborhood  
\citep{1962AJ.....67..486V,1963ApJ...137..758S,1975VA.....19..299L} -- 
see Fig.~\ref{fig:mdwarf}. This {\em problem} has deserved careful attention in the 
literature, with solutions going from variations of the IMF 
\citep{1996RMxAA..32..179C,2000AA...353..269M} 
to inhomogeneous chemical evolution and star formation 
\citep{1993ApJ...413..633M}. Among them, a continuous metal-poor gas inflow 
sustaining the formation of the G stars seems be the preferred mechanism 
\citep{2001coev.conf..223P,2005AG....46d..12E}. The explanation was first 
proposed by \citet{1972NPhS..236....7L}, who discovered that the 
SF maintained by constant metal-poor gas accretion reaches 
a constant value set only by the stellar yield  (see Eq.~[\ref{eq:metallicity}]), 
which implies a value around the solar metallicity. In the context of this explanation, 
the apparent deficit of sub-solar metallicity G dwarfs is actually an excess of solar metallicity G dwarf 
stars formed over time out of an ISM always near equilibrium at approximately the solar metallicity.

The G dwarf problem has also been observed in K dwarfs \citep[e.g.,][]{2004AA...419..181C} and in
M dwarfs \citep[e.g.,][]{2012MNRAS.422.1489W}, and it exists in other galaxies as well  
\citep[e.g.,][]{1996AJ....112..948W}. Current chemical evolution models resort to 
metal-poor gas inflow to reproduce the spatial distribution of stellar metallicities 
observed in the disk of spirals 
\citep[e.g.,][]{2001ApJ...554.1044C,2010AA...512A..63M,2016MNRAS.462.1329M,2016MNRAS.455.2308P}.
Such gas inflow is needed for the same reasons invoked to solve the G dwarf problem.

\subsection{Existence of a minimum metallicity for the star-forming gas} 

\begin{figure}
\begin{center}
\includegraphics[width=0.7\textwidth]{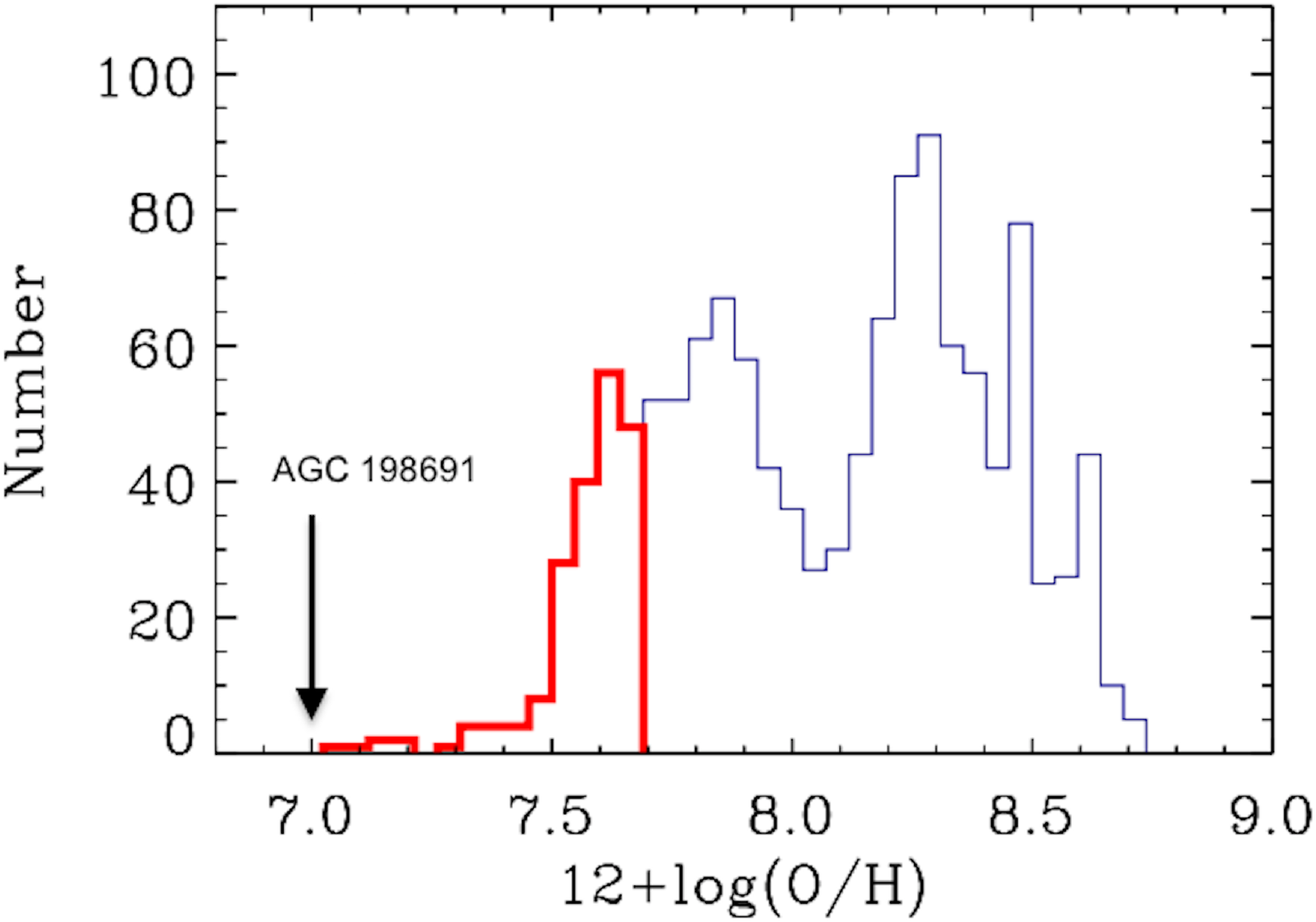}
\end{center}
\caption{Distribution of oxygen abundance [$12+\log({\rm O/H})$]
for all the objects 
found in the search for low-metallicity galaxies by 
\citet{2016ApJ...819..110S}. The solar metallicity is set at 
$12+\log({\rm O/H})_\odot=8.69$, therefore,  the value $12+\log({\rm O/H})=7.69$ 
separates the XMPs (the thick  red line) and the failed XMP candidates 
(the thin blue line). There is a clear drop in the distribution towards 
low metallicity, with no object metal-poorer than 2\,\%\ of the solar metallicity.
AGC~198691 is the star-forming galaxy with the lowest metallicity known 
\citep{2016ApJ...822..108H}, and the lower limit it sets is marked 
with an arrow. 
Figure adapted from \citet{2016ApJ...819..110S}. 
}
\label{fig:histogram}
\end{figure}
XMPs are defined as galaxies where the gas that produces stars has a metallicity
smaller than 10\,\%\ of the solar metallicity. They turn out to be quite rare. 
Systematic searches, such as that carried by \citet{2016ApJ...819..110S},
render a few hundred objects in catalogs containing of the order of one million galaxies 
(XMP represent $\ll 0.1\,$\%\ of the known galaxies). Interestingly, their metallicity, and
therefore the metallicity of all local galaxies, seems to have 
a lower limit at around 2\%\ of the  the solar metallicity. Figure~\ref{fig:histogram} displays
the distribution of metallicities found by \citet{2016ApJ...819..110S} 
in the search
for XMP candidates in SDSS. It has a sharp cut-off at low metallicity, with no galaxy  with 
$12+\log({\rm O/H}) \le 7.0$.
The current record-breaking object is AGC~198691, with a metallicity around 2.1\,\%\ times
the  solar metallicity \citep{2016ApJ...822..108H}. It is also included in Fig.~\ref{fig:histogram}
for reference.

The existence of this metallicity threshold  
is not an artifact. Observers have been looking for record-breaking galaxies during 
the last 45 years, after the discovery of the prototypical XMP galaxy IZw~18 
\citep{1970ApJ...162L.155S}. These efforts led to enlarging the number of known XMPs
\citep{1991AAS...91..285T,2009AA...505...63G,2015AA...579A..11G,
2011ApJ...743...77M,2012AA...546A.122I,2016ApJ...819..110S}, 
but the lower limit metallicity set by  IZw~18 
remains almost unchanged \citep[the metallicity of  IZw~18 is about 3\,\%\ times
the solar metallicity; see, e.g.,][]{2005ApJS..161..240T}.

Several explanations have been put forward to account for the existence of 
a minimum metallicity in the gas that forms stars.  \citet{1986ApJ...300..496K} point out the 
self-enrichment of the H\,II region used for measuring. However, the time-scale for the 
SN ejecta to cool down and mix is of the order of several hundred Myr \citep[e.g.,][]{2001ApJ...560..630L}
and, thus, longer than the age of the H\,II regions. This fact eliminates the possibility of self-contamination. 
Self-enrichment by massive star winds seems to be negligible too \citep{2006AA...450L...5K}. 
\citet{2011EAS....48...95K} suggest the pre-enrichment of the proto-galactic clouds,
but it is unclear why there should be a minimum metallicity for 
such clouds, except perhaps the value set by the metal contamination 
produced by Pop~III stars. Pop~III star contamination has been suggested too 
\citep[][]{1995ApJ...451L..49A,2005ApJS..161..240T}, but the expected level of metal enrichment
is of the order of $10^{-4}$ times the solar metallicity \citep{2004ARAA..42...79B,2013ApJ...772..106M}, 
and so much lower than the observed threshold.  

Alternatively, if the SF is feeding from gas of the IGM, there is a minimum gas-phase
metallicity set by the metallicity of the local IGM. This possibility   
provides a natural explanation for the long-lasting puzzle.
Numerical simulations predict the local cosmic web gas to have a metallicity of the 
order of 1\,\% times the solar value  
\citep[e.g.,][]{2012MNRAS.423.2991V,2016MNRAS.459..310R}.
The metal content of the IGM has been rising over time contaminated by  
galactic winds, and now it happens to be at the level of the observed  
metallicity threshold.

Other independent observations also support the existence of a minimum 
metallicity in the CGM of galaxies at the level of 1\,\% \ times the solar metallicity.
\citet{2013ApJ...770..138L} measure the distribution of metallicity 
of Ly$\alpha$ absorbing clouds around galaxies with redshift up to one.
The clouds are observed  in absorption against background QSOs. The distribution of metallicity
turns out to be bimodal with typical values around 2.5\,\%\ solar and 50\,\%\ solar.
The high-metallicity branch is expected to represent galaxy outflows, whereas
the low-metallicity branch corresponds to inflows. The lowest measured metallicity 
turns out to be 1\,\% times the solar metallicity.
In a completely independent type of work, \citet{2013AA...558A..18F} studied
the H\,I gas around XMP galaxies. The XMPs happen to be extremely
gas-rich, with gas fractions typically in excess of 10. Assuming that all the metals
produced by the observed stellar populations have been diluted in their huge H\,I reservoirs,
the metallicity of the H\,I is again around a few percent of the solar value.  
The same conclusion has been recently reached by 
\citet{2016MNRAS.tmp.1366T}.
Sometimes the metallicity of the H\,I gas can be measured directly using 
UV lines in absorption against the stellar light of the galaxy. In the case 
of IZw~18, \citet{2013AA...553A..16L} find that H\,I region abundances are also
around 1\,\%\ of the solar value. 


\subsection{Origin of $\alpha$-enhanced gas forming stars in local galaxies}\label{sec:alphae}

\begin{figure}
\begin{center}
\includegraphics[width=0.7\textwidth]{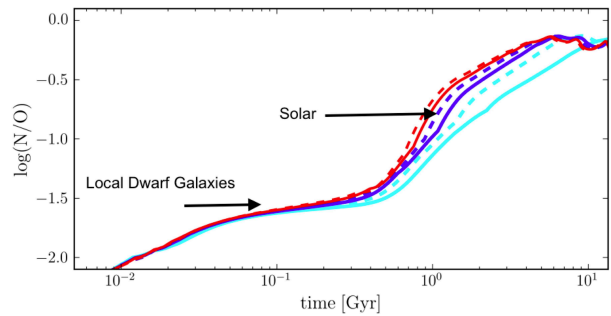}
\end{center}
\caption{
Time evolution of N/O for chemical evolution models aimed at reproducing the 
relationship between N/O and O/H exhibited by the galaxies of the local universe. 
The different curves correspond to different $\tau_g$ 
going from 2\,Gyr (solid magenta line) to 0.3\,Gyr  (red dashed line).
The figure includes the N/O found in the Sun and the characteristic 
value observed in metal-poor galaxies of the local universe. 
By the time that the burst is 2\,Gyr old, N/O has already reached the solar value
in all models.
Adapted from \citet{2016MNRAS.458.3466V}.}
\label{fig:alphae}
\end{figure}

The gas forming stars in some of the local dwarf galaxies has elemental abundances which do not
scale with the solar abundances.  The gas is {\em $\alpha-$enhanced},  using the terminology 
employed when studying 
stellar  populations in massive galaxies and in the MW halo
\citep[e.g.,][]{2011AA...535L..11A,2015MNRAS.449.1177V}.
The star-forming gas often shows $\log({\rm N/O})\simeq -1.5$
\citep[e.g.,][]{2015MNRAS.448.2687J,2016MNRAS.458.3466V} which
is much smaller than the solar value \citep[of the order of -0.86;][]{2009ARAA..47..481A}.
Stars and gas with $\alpha-$enhanced composition
are expected to be formed from the ejecta of young stellar populations 
\citep[e.g.,][]{1999ApJ...511..639I}, so that low mass stars have not 
had the time to explode as SN, and so, elements like Fe are 
underrepresented compared to the solar composition.  N is one of these elements, even 
though N is also produced in intermediate-mass  stars \citep[e.g.,][]{2000ApJ...541..660H}. 
Figure~\ref{fig:alphae} shows the time evolution of N/O in several different model starbursts 
with very   different  SF efficiencies  (i.e., with different $\tau_g$ in the 
parlance used in Sect.~\ref{sec:intro}). 
\citet{2016MNRAS.458.3466V} compute them to model the 
relationship between N/O and O/H observed in the galaxies of the local universe.
Independently of the SF efficiency, when a starburst  is 2~Gyr old it has already 
reached ${\rm N/O} \simeq {\rm N/O}_\odot$ (see Fig.~\ref{fig:alphae}).  Something equivalent is 
shown by \citet[][]{2005AA...434..531K} when modeling the evolution of systems that accrete 
large amounts of metal-free gas. After a transient phase that lasts around 2\,Gyr, 
the  system returns to the original ${\rm N/O}_\odot$.  
The fact that the gas is $\alpha$-enhanced is consistent with the thorough study
on the heavy element abundances in local star-forming galaxies carried out by \citet{1999ApJ...511..639I}.
The $\alpha$-elements Ne, Si, S and Ar, produced by SN explosions of massive ($> 10 M_\odot$)
stars that also produce O, show the same $\alpha$-element/O ratio as the Sun.  However, 
O turns out to be overabundant with respect to N, C and Fe by factors of around three, which are
typical of MW halo stars.  

It is well known that all galaxies, including the local dwarf galaxies, 
have their mass dominated by old stars with ages extending all 
the way back to the origin of the universe 
\citep[e.g.,][see also Fig.~\ref{fig:sfr_time}]{2004Natur.428..625H,2012ApJ...756..163S}. 
The fact that N/O is low in the star-forming gas of some local dwarfs  implies that their
evolved stellar populations are not the source of the  metals. 
If the metals are not produced by the observed stellar populations, 
where do they come from?   They likely come  together with the accreted
gas from the IGM. The IGM gas is expected to have $\alpha$-enhanced chemical  composition 
since the metals its contains were produced by dwarfs galaxies in 
the early universe when the stellar populations were 
young \citep[e.g.,][]{2005ApJ...629..636A,2012MNRAS.423.2991V,2013MNRAS.435.3500Y}. 

The presence of   $\alpha$-enhanced gas forming stars at high redshift seems to be common.
\citet{2016ApJ...826..159S} argue that the unusually
high excitation of the gas forming stars in galaxies at redshift around 2
is due to their extreme $\alpha-$enhanced composition, with O/Fe$\sim 5$ times
the solar ratio. Such extreme conditions yield Fe-poor stars in an ISM of moderate-high  
metallicity as traced by O.  Similar physical conditions in the local universe may be responsible for 
the presence of high excitation narrow He\,II lines in the spectra of metal-poor
galaxies of the local universe \citep{2012MNRAS.421.1043S}.


\subsection{The metallicity of the quiescent BCD galaxies}

The blue-compact dwarf (BCD) galaxies are going through an intense starburst phase that cannot be 
maintained for long.   Consequently, there must be many dormant galaxies in a pre or a post BCD phase. 
They are called  quiescent BCDs (QBCD). BCDs possess one or a few bright starburst 
regions on a low-surface host. Masking out the high-surface knots,   
\citet{2007AA...467..541A,2009AA...501...75A}  characterized  the photometric 
properties of the host galaxy, which likely corresponds to the underlying QBCD. Using these properties as
reference, \citet{2008ApJ...685..194S} found out that the SDSS catalog contains as many as
30 QBCDs per BCD.  BCDs and QBCDs seem to be drawn from a single galaxy  
population that metamorphose with a cycle where the quiescent phase lasts 30 times longer 
than the  star-forming phase.  However, this interpretation presents a difficulty. 
The gas metallicity of the QBCDs is systematically higher than the metallicity of the BCDs,
which cannot happen if the transformation between QBCD and 
BCD occurs through closed-box evolution, where the precursor always
has lower metallicity than the offspring. The problem naturally goes away if the BCD 
phase is triggered by the accretion of external metal-poor gas that feeds the observed SF
episode. The external driving of the BCD phase also explains why the stellar metallicity 
of BCDs and  QBCDs agrees, even though their gas metallicity differs \citep{2009ApJ...698.1497S}.
The stellar populations of BCDs and QBCDs are statistically the same because only a small fraction of 
the galaxy stellar mass is built up in each new burst. 

\subsection{Direct measurement of inflows in star-forming galaxies}

To the best of our knowledge, there is no direct measurement of metal-poor 
gas inflows in the CGM of galaxies. Finding whether a particular Doppler shift
corresponds to inflows or outflows is tricky because the same 
signal is provided by inflows or outflows depending on whether 
the source is in the  foreground or the  background.  
Fortunately, the sense of motion can be disambiguated when the gas is absorbing
or emitting  against the spectrum of the galaxy. In this cases the gas is in the foreground,
which breaks the degeneracy.

Thus, gas inflows have been detected in absorption against stellar spectra 
\citep[e.g.,][]{2016AAS...22731210R},
but they are probably associated with metal rich gas returning to the galaxy after being 
ejected by a starburst or an AGN 
\citep[e.g.,][]{2008AARv..15..189S,2012MNRAS.419.1107M,2014IAUS..298..228F}. 
Neutral gas has been studied in absorption in several metal-poor objects
\citep[][]{2003ApJ...595..760A,2005ApJ...618..247C,2009AA...494..915L,2013AA...553A..16L}.
However, the relative velocity between the gas absorption and the systemic velocity of the galaxy 
is too  small to tell whether the gas is falling in or going out.  
If anything, the gas seems to be flowing out with a mild velocity between 10 and 20 km~s$^{-1}$ 
\citep{2013AA...553A..16L}.  The origin of this putative metal-poor outflow is puzzling.    

Recent observations by \citet{2016MNRAS.461.1816F} show signs of infall in DLAs
eclipsing QSOs, i.e., dense H\,I gas so close in redshift to the QSO that  it 
blocks the  Ly$\alpha$ emission of the source.  Their metallicities are relatively low, and  
most of them have redshifts larger than that of their background QSO, strengthening the idea 
that they could be associated with some low metallicity infalling
material accreting onto the QSO host galaxies.

Even though the works by \citet{2012ApJ...760L...7K,2015ApJ...815...22K} 
do not have kinematical information, they are very suggestive of gas 
inflows in star-forming galaxies. They study the azimuthal distribution 
of Mg II and O VI absorbers around the galaxy responsible for the absorption. These absorption
systems trace gas in the CGM and the IGM, depending on the distance 
to the galaxy. \citet{2012ApJ...760L...7K,2015ApJ...815...22K} find that the 
absorption preferentially occurs along the directions pointed out by the minor and the 
major axes of  the central galaxy. The absorbers aligned with the 
major axis are expected to show gas inflows, with the absorbers in the direction of 
the minor axis corresponding to gas outflows. 
In order to secure the whole scenario, one would need to have measurements of the
velocity and metallicity of the Mg\,II systems. Unfortunately, they are not available. 
However, two independent measurements support the above 
interpretation. Firstly, outflows prefer the direction perpendicular to the plane 
of the galaxy because Mg\,II outflows are faster in face-on galaxies 
\citep[][]{2014ApJ...794..130B,2014ApJ...794..156R}. 
Secondly, the metallicity of Lyman limit systems (i.e., gas 
clumps of moderate H\,I column density) is observed to be bimodal, 
with one peak at low metallicity and the other at high metallicity \citep{2013ApJ...770..138L}.
This bimodality is to be expected if the observed absorption is produced by metal-poor inflows and
 metal-rich outflows.


\section{Obvious complications and future trends}\label{sec:complications}

The evidence discussed in Sect.~\ref{sec:evidence} is all indirect. 
We have to rely on models  to interpret the observables  as evidence for gas accretion. 
The toy model setting up the scene in Sect.~\ref{sec:intro}, and then
used in many of the above arguments, oversimplifies many aspects of the accretion 
process that are important to identify in real galaxies the observational signatures of 
ongoing cosmological  gas accretion. 
The purpose of this section is to point out some of the obvious complications of 
a more realistic modeling, and also to point out new observational pathways that may reveal
connections between SF and the gas of the IGM.

{\bf The CGM and the galactic fountain.} The CGM is a complex region where gas 
ejected  from the galaxy and gas falling into the galaxy coexist. 
The gas inflow is not only of cosmological origin. Part of the 
metal-rich materials ejected in previous SF episodes fall back
to the disk  in what is called galactic fountain 
\citep[e.g.,][]{2006MNRAS.366..449F,2008MNRAS.386..935F,2015MNRAS.452.3593H}. 
The mixing with metal-rich gas from SF processes  speeds up the cooling of the hot metal-poor CGM, 
and the fountain also returns gas that was never ejected  
\citep{2008MNRAS.388..573M,2010MNRAS.404.1464M,2011AA...525A.134M,2012MNRAS.419.1107M}.

The structure of the CGM is extremely complicated  according to the current numerical simulations. 
The morphology of the gas streams  
becomes increasingly complex at higher resolution, with large coherent flows revealing density and temperature 
structure at progressively smaller scales, and with no evidence that the substructure is properly captured
in the simulations \citep{2016MNRAS.460.2881N}.  Multiple gas components co-exist 
at the same radius within the halo, making radially averaged analyses misleading. 
This is particularly true where the hot quasi-static halo interacts with cold rapidly-inflowing IGM accretion. 
Some of the resulting complications are revealed in the study of metal-poor DLAs carried out by  
\citet{2016MNRAS.457..487Y}.  The majority of the metal-poor DLAs are far from the central galaxy 
($\geq 20\,$kpc)  and result from the cold gas streams from the IGM. 
In the migration inwards to the galaxy, they mix up with high-metallicity gas from 
stellar outflows,  removing themselves from the metal-poor category.
The change from metal-poor to metal-rich  complicates the observational identification of the
gas coming from the IGM.  IGM gas clouds that get mixed with the CGM and become metal-rich
are also found in the simulation by \citet{2014ApJ...795...99G}.
%
%
The difficulties of interpretation are also put forward by \citet{2014MNRAS.444.1260F}.
Using cosmological simulations, \citeauthor{2014MNRAS.444.1260F}  examine how H\,I and metal 
absorption lines trace the dynamical state of the CGM around low-redshift galaxies.
Recycled wind material is preferentially found close to galaxies, and is more dominant in low-mass 
halos.  Typical H\,I absorbers trace unenriched ambient material that is not participating in 
the baryon cycle,  but stronger H\,I absorbers arise in cool, metal-enriched inflowing gas.  
Instantaneous radial velocity measurements are generally poor at distinguishing between 
inflowing and outflowing gas, except in the case of very recent outflows.

Galactic fountains and metal-rich gas produced in previous 
starbursts are very important because their presence complicates 
the search for metal-poor gas  inflows in the CGM of galaxies.  However, their role in 
sustaining SF should not be overestimated. Most of the gas used to produce 
stars at any time is pristine. It was never pre-processed by a star. 
This issue has been recently quantified by \citet{2016MNRAS.456.1235S}
using galaxies from the EAGLE numerical simulation \citep{2015MNRAS.446..521S}.
For MW-like galaxies, recycled stellar ejecta account for only 35\,\%\ of the SFR 
and 20\,\%\ of the stellar mass.  The contribution was even less important in the past.
The toy model in Sect.~\ref{sec:intro} provides the right order of magnitude  
for this estimate \citep[see ][]{2014AARv..22...71S}.

{\bf Star formation generated by gas accretion.}
The current cosmological numerical simulations produce model galaxies that
look impressively realistic, and follow most of the well known 
scaling relations \citep[e.g.,][]{2014MNRAS.444.1518V,2015MNRAS.446..521S}.
They provide the theoretical framework to understand the formation of galaxies 
and the role played by cosmological gas accretion in maintaining SF.
However, their limited resolution and the dependence of many 
predictions on the adopted sub-grid physics make them less reliable to study
how individual starbursts grow out of the gas that arrives 
to the galaxy disk.  (Predictions on individual 
star-formation events are discussed in Sec.~\ref{sec:lopsided}.)
Improving this aspect is critically important to secure the interpretation of many 
observables currently used as evidence for cosmological gas accretion.  

{\bf Imaging the cosmic web.} 
Much of our knowledge on the CGM and IGM comes from observing absorption lines against 
background sources that happen to be next to galaxies.  However, the observation and analysis of these
absorption systems is extremely time-consuming, and even the best cases only provide 
a very  sparse sampling of the CGM and IGM around individual galaxies.  
A complementary approach is observing the cosmic web gas in emission.  
The mechanisms to produce such emission are varied. Ly$\alpha$ (as well as H$\alpha$)
can be produced by electron collisions within a gas stream that releases the gravitational energy 
gained as gas flows from the  IGM into the galaxy halo 
\citep{2009MNRAS.400.1109D,2010MNRAS.407..613G,2010ApJ...725..633F}.
Emission also results from fluorescence induced by an intense UV radiation field 
such as that produced by a large nearby starburst or a QSO 
\citep[e.g.,][]{1987MNRAS.225P...1H,2012MNRAS.425.1992C,2015AA...581A.132A}.

Radio emission is also expected to trace the cosmic web. In this sense,
the search for {\em dark galaxies} (i.e., objects emitting in
21\,cm without optical counterpart)  is a very revealing and active field of research 
\citep[e.g.,][]{2014ApJ...787L...1C,2015MNRAS.452.2680S,2015ApJ...811...35J,2015ApJ...801...96J}.
Gas filaments associated with star-forming galaxies are very interesting too
\citep[][]{2008AARv..15..189S,2012AA...537A..72L,2013AA...558A..18F}. 
Radio data easily provide kinematical information, which is so
important when investigating flows. In this context, a large filament of molecular gas
accreting onto a group of massive high redshift galaxies has been recently discovered by Ginolfi et al. 
(2016, private communication). This observation is intriguing, but it may open up a new way of 
addressing the search for IGM gas.   
 
Extremely promising is the recent discovery of an extended Ly$\alpha$ blob connected to a QSO
\citep{2014Natur.506...63C}. The emission extends beyond the virial radius of the host galaxy so that
it traces gas in the IGM. The existence of extended Ly$\alpha$ emission around QSOs seems to be very
 common when the observation is deep enough \citep{2016arXiv160501422B}. The 
sensitivity is very much improved using spectrographic observations, which  
also have the capability of providing the eagerly needed kinematical information (see the chapter by 
Cantalupo in this Book).  Extended Ly$\alpha$ halos are common around all kinds of galaxies 
\citep[e.g.,][]{2016MNRAS.455.3991R,2016MNRAS.457.2318M}.


\section{Conclusions}\label{sec:conclusions}

Cosmological numerical simulations predict that the SF in regular 
disk galaxies is feeding from metal-poor gas accreted from the cosmic web. 
The observational evidence for a relation between SFR  and external gas accretion is both 
numerous and indirect   (Sect.~\ref{sec:evidence}). One necessarily has to rely on modeling and 
analysis to identify the existing hints as  actual signs of  SF driven by metal-poor gas accretion. 
Thus, the comparison with numerical simulations and a meticulous interpretation of the 
observations turn out to be mandatory. There is no reason to believe that the magnitude 
of the problem will change significantly in the near future.  
The IGM gas is predicted to be tenuous and  ionized, and so 
extremely elusive observationally.
The IGM gas gets mixed with metal-rich outflows, and may loose its distinctive metal-poor 
character to complicate the study. In addition,  metal-rich recycled 
material is also re-accreted, and often velocities are useless to separate inflows 
from outflows.
Although things have much improved during the last few years, 
theoretical predictions are still very unspecific as far as
the details is concerned. These details are needed
to interpret particular observational results as evidence 
for accretion  (Sect.~\ref{sec:drops}). 
Finally, the difficulty to distinguish gas accretion events from 
gas-rich minor mergers confuses the interpretation even further (Sect.~\ref{sec:lopsided}).


%

\begin{table}
\caption{List of acronyms and symbols defined and used along the text}
\label{acronyms}      
\begin{tabular}{ll}
\hline\noalign{\smallskip}
Acronym & Expansion \\
\noalign{\smallskip}\hline\noalign{\smallskip}
ADS & NASA Astronomical Data System \\
BCD &Blue compact dwarf\\
CGM &Circum-galactic medium\\
DLA&Damped Lyman-$\alpha$ absorbers\\
EAGLE& Evolution and Assembly of\\ 
&GaLaxies and their Environments\\
&\citet{2015MNRAS.446..521S}\\
FMR&Fundamental metallicity relation\\
HCM & HII-Chi-Mistry \citep{2014MNRAS.441.2663P}\\
IGM&Inter-galactic medium\\
ISM&Interstellar medium\\
IMF&Initial mass function\\
MZR& (Stellar) Mass metallicity relation\\
MW&Milky Way\\
\noalign{\smallskip}\hline
\end{tabular}
\begin{tabular}{|ll}
\hline\noalign{\smallskip}
Acronym & Expansion \\
\noalign{\smallskip}\hline\noalign{\smallskip}
${M}_\star$& Stellar mass\\
$M_g$& Gas mass\\
QBCD &Quiescent blue compact dwarf\\
QSO&Quasar\\
SDSS& Sloan Digital Sky Survey\\
SF&Star formation\\
SFR&Star formation rate\\
SN, SNe& Supernova, Supernovae\\
SPH&Smoothed particle hydrodynamics\\
UV&Ultraviolet\\
$w$&Mass loading factor (Eq.~[\ref{eq:mass_loading}])\\
XMP& Extremely metal poor\\
$Z$ &Metallicity of the gas (Eq.~[\ref{eq:metallicity}])\\
&\\
\noalign{\smallskip}\hline
\end{tabular}
\end{table}

\acknowledgement
Special thanks are due to Casiana Mun\~oz-Tu\~n\'on,  Debra Elmegreen, and  Bruce Elmegreen,
for continuous support and for long thoughtful discussions on almost every topic included in the 
work. 
I am also indebted to Mercedes Filho, who pointed out many of the references cited in the 
work, and shared with me her expertise on H\,I.  
The interpretative aspects of the work own much to the collaboration with 
Claudio Dalla Vecchia and Daniel Ceverino. They were always willing discuss
the physical aspects of the accretion and the star-formation process.  
Thanks are due to R. Amor\'\i n for the discussions that led to the
writing of Sect.~\ref{sec:alphae}. 
Thanks are also due to the editors of the Book for giving me the opportunity to 
contribute, and for their patience to have my contribution finished.
This work has been partly funded by the Spanish Ministry of Economy and 
Competitiveness, project {\em Estallidos} AYA2013--47742--C04--02--P. 
%

%
\newcommand\aj{AJ}
\newcommand\apj{ApJ}
\newcommand\apjl{ApJ}
\newcommand\apjs{ApJS}
\newcommand\mnras{MNRAS}
\newcommand\aapr{A\&ARev}
\newcommand\araa{ARA\&A}
\newcommand\aap{A\&A}
\newcommand\aaps{A\&AS}
\newcommand\nat{Nature}
\newcommand\pasp{PASP}
\newcommand\nar{NewARev}
\newcommand\na{NewA}
\newcommand\pra{PhyRevA}
\newcommand\ap{$\approx$}
\newcommand\jcap{JCAP}
\newcommand\rmxaa{RMxAA}
\newcommand\fcp{FCPh}
\newcommand\pasa{PASA}
\newcommand\pasj{PASJ}

\end{document}